\documentclass[aps,prb,preprint,superscriptaddress]{revtex4}

\usepackage{epsfig} 

\begin{document}

\title{Magnetic excitations in ferro-pnictide materials
controlled by a quantum critical point into hidden order}




\author{J.P. Rodriguez}

\affiliation{Department of Physics and Astronomy, 
California State University, Los Angeles, California 90032}

\date{\today}

\begin{abstract}
The two-orbital $J_1$-$J_2$ model that describes a square lattice of frustrated
spin-1 iron atoms is analyzed within the 
linear spin-wave approximation and by exact diagonalization over a $4 \times 4$ cluster. 
A quantum critical point (QCP) is identified that separates hidden magnetic
order at weak Hund's rule coupling
from a commensurate spin density wave (cSDW) at strong Hund's rule coupling.  
Although the moment for cSDW order is small at the QCP,
the critical linear spin-wave spectrum shows strong low-energy excitations
centered at the wavenumbers that correspond to cSDW order.
These disperse anisotropically.  
A fit to the magnetic excitation spectrum of ferro-pnictide materials
obtained recently by inelastic neutron scattering measurements 
notably accounts for the absence of softening
at the wavenumber that corresponds to N\'eel order.

\end{abstract}

\maketitle

\section{Introduction}
The discovery of ferro-pnictide superconductors has set a new and
unforseen direction in the search for high-temperature superconductivity.\cite{new_sc}
A common structural feature of these materials are 
square lattices of iron atoms that are stacked up on top of each other.\cite{delacruz}
Density functional theory (DFT) calculations find that the iron 3d orbitals contribute most
of the electronic spectral weight near the Fermi surface.\cite{haule_08}$^{,}$\cite{cao_08}
This implies that the iron $3d$ electrons surprisingly play an active role in the superconductivity.
In contrast, DFT calculations also predict that
the iron moments order antiferromagnetically along the $a$ and along the $c$ axes
of the iron lattice at low temperature in the undoped parent compounds,\cite{dong_08}
with an ordered moment of approximately $2$ Bohr magnetons ($\mu_B$).\cite{ma_08}
Elastic neutron diffraction studies have confirmed the prediction of long-range
commensurate spin-density wave (cSDW) order, 
but they find a much lower ordered moment that can be a small
fraction of the Bohr magneton ($0.36\,\mu_B$) in some cases.\cite{delacruz}
It was suggested early on that Heisenberg exchange between both nearest-neighbor ($J_1$)
and next-nearest neighbor ($J_2$) spins on a square lattice of iron atoms 
could account for the low
ordered moment that is observed because of magnetic frustration.\cite{Si&A}
Recent calculations based on the linear spin-wave approximation\cite{j1j2_sw}
indicate that this effect requires fine tuning however,\cite{thalmeier_10}
and that it occurs only very close to the point of maximum frustration, 
$J_2 = 0.5 \, J_1$.
A possibly more viable explanation is 
that the low ordered cSDW moment commonly observed
in ferro-pnictide parent compounds is due to proximity to hidden magnetic order
among particular iron $3d$ orbitals such that the magnetic moments cancel out per iron atom.
The author and Rezayi have shown that such violations of Hund's rule
can occur in multi-orbital Heisenberg models with off-diagonal frustration.\cite{jpr_ehr_09}

Spin-wave excitations of the cSDW groundstate
have also been detected in parent compounds to ferro-pnictide
superconductors at low temperature
by inelastic neutron scattering.\cite{zhao_08}$^,$\cite{diallo_09}
Contrary to the prediction of 
the conventional $J_1$-$J_2$ model over the square lattice however,\cite{oitmaa_10}
the measured spin-wave spectrum fails to soften at the wave number associated with N\'eel order,
$(\pi / a, \pi / a)$. Here $a$ denotes the square lattice constant.
It shows a local maximum there instead.\cite{zhao_09}
Fits to the measured spin-wave dispersion invoke 
unphysically large anisotropy in the Heisenberg
spin exchange across nearest-neighbor links on the square lattice of iron atoms
in order to account for the dispersion at the Brillouin zone boundary.
More recently, low-energy magnetic  excitations centered at the cSDW wave numbers
have been detected in a ferro-pnictide {\it superconductor}.\cite{hayden_10}
These disperse up in energy in an anisotropic fashion
that is  consistent with the presence of  some degree of cSDW order.

In this paper, we continue to explore the hidden magnetic order that is
predicted to exist in multi-orbital Heisenberg models over the square lattice when
the diagonal and the off-diagonal exchange coupling constants
are sufficiently different.\cite{jpr_ehr_09}
Two spin-1/2 orbitals per iron atom are assumed, 
which implies spin-1 iron moments if Hund's rule is obeyed.
A quantum critical point\cite{sachdev_qcp}
is identified within the linear spin-wave approximation
that separates hidden ferromagnetic or hidden N\'eel order
at weak Hund's rule coupling from a cSDW at strong Hund's rule coupling.
Numerical exact diagonalizations of the two-orbital Heisenberg model 
over a $4 \times 4$ lattice indicate
that the moment associated with cSDW order decays strongly as it enters hidden magnetic order
through the putative quantum critical point.
The linear spin-wave approximation, on the other hand, 
finds observable low-energy spin-wave excitations at the quantum critical point
that disperse linearly from the cSDW wave numbers in an anisotropic fashion.
A spin gap begins to grow there upon entering hidden magnetic order.
Observable low-energy spin-wave excitations due to hidden order
that are centered at zero momentum
are also predicted by the linear spin-wave approximation,
but these have very small spectral weight.
Fits of the spin-wave spectrum observed by inelastic neutron scattering in 
ferro-pnictide materials\cite{zhao_09}$^,$\cite{hayden_10}
to the critical linear spin-wave spectrum are achieved.
They notably account for 
the dispersion of the low-energy spinwaves centered at the cSDW wave numbers
and for the local maximum that exists at the wavenumber associated with N\'eel order.
A critical Hund's rule coupling of order $100$ meV is extracted from these fits.
Below we discuss useful theoretical aspects of the 
Heisenberg model for ferro-pnictide materials\cite{Si&A}
before we go on to analyze it.

\section {Multi-Orbital Heisenberg Model}
Recent low-temperature measurements of the optical conductivity 
in the cSDW phase of parent compounds to ferro-pnictide high-$T_c$ superconductors 
find a strong suppression of the integrated kinetic energy for the conduction electrons
compared to that predicted by electronic band-structure calculations.\cite{basov_09}
A low carrier density 
in the cSDW phase of ferro-pnictide materials
is consistent with
quantum-oscillation experiments\cite{sebastian_08}$^,$\cite{analytis_09}
and with angle-resolved photoemission studies\cite{richard_10} 
that  find evidence for tiny 2D Fermi-surface ``spots'', 
each on the order of 1\% of the Brillouin zone.
Also,
the dependence on temperature shown by the electrical resistance
of such  parent compounds 
indicates proximity to a metal-insulator transition.\cite{wang}  
Undoped ferro-pnictide materials in the cSDW phase
then may very well lie near a Mott transition,
where conduction electrons experience moderately strong repulsive  interactions.
By continuity
with the limit of strong inter-electron repulsion,
we therefore believe that a local-moment description of magnetism in parent compounds
to iron-based high-$T_c$ superconductors is a justifiable starting point at low temperature.
The absence of a Stoner continuum of incoherent magnetic excitations 
reported in recent experimental studies of magnetic excitations
in the cSDW phase of a ferro-pnictide material
is  consistent with this view.\cite{zhao_09}
Consider then the following
multi-orbital spin-1/2 Hamiltonian that 
contains near-neighbor Heisenberg exchange among local iron moments
within isolated layers in addition to Hund's-rule coupling:\cite{Si&A}
\begin{equation}
H = {1\over 2} J_0 \sum_i \biggl[\sum_{\alpha} {\bf S}_i (\alpha)\biggr]^2 +
    \sum_{\langle i,j \rangle} \sum_{\alpha , \beta} J_1^{\alpha,\beta}
                                      {\bf S}_i (\alpha) \cdot {\bf S}_j (\beta) +
    \sum_{\langle\langle i,j \rangle\rangle} \sum_{\alpha , \beta} J_2^{\alpha,\beta}
                                      {\bf S}_i (\alpha) \cdot {\bf S}_j (\beta).
\label{j0j1j2}
\end{equation}
Above, ${\bf S}_i(\alpha)$ is the spin operator that acts on the spin-1/2 state of orbital
$\alpha$ in the iron atom at site $i$.  The latter runs over the square lattice
of iron atoms that make up an isolated layer.  
The application of Hund's rule is controlled by a local ferromagnetic
Heisenberg exchange constant $J_0 < 0$, 
while nearest neighbor and next-nearest neighbor Heisenberg exchange
across the links $\langle i,j\rangle$ and $\langle\langle i,j\rangle\rangle$
is controlled by the tensor exchange constants $J_1^{\alpha,\beta}$ and $J_2^{\alpha,\beta}$, respectively.
The strength of the crystal field at each iron atom
determines the number of orbitals per iron atom above.
It can be as low as two for strong crystal fields compared to the Hund's-rule coupling, 
in which case spin-1/2 moments exist on 
the iron $3d_{xz}$ and $3d_{yz}$ orbitals.\cite{cao_08}$^,$\cite{li_08}
Four $3d$ orbitals per iron atom carry spin-1/2 moments 
in the case of weak crystal fields compared to the Hund's-rule coupling,
on the other hand.\cite{haule_08}$^,$\cite{Si&A} 
Last, the above model Hamiltonian reduces to the conventional
$J_1$-$J_2$ model over the square lattice 
when Hund's rule is enforced,\cite{j1j2_sw}$^,$\cite{thalmeier_10}$^,$\cite{oitmaa_10}
 $-J_0 \rightarrow \infty$,
where the spin at each site is given by half the number of orbitals, $S = 1$ or $S = 2$.

We now illustrate a useful transformation of the model Hamiltonians (\ref{j0j1j2})
over the square lattice
that relates eigenstates and eigenvalues with different nearest-neighbor Heisenberg exchange
coupling constants.
Let $p$ denote a permutation of the orbitals.  Consider then the partial ``reflection''
$P_{A(B)}$:
\begin{equation}
\quad {\rm orbital}\quad \alpha \rightarrow p(\alpha) 
\quad {\rm on\ \ the\ }A(B)\ {\rm sublattice}
\label{trans_orb}
\end{equation}
of the N\'eel state.
If $\Psi$ is an eigenstate of the original model Hamiltonian (\ref{j0j1j2})
with energy $E$,
the transformed state $\Psi^{\prime} = P_{A(B)} \Psi$ is clearly then 
an eigenstate of the transformed Hamiltonian $H^{\prime} =  P_{A(B)} H P_{A(B)}^{-1}$
with energy $E$.
Importantly, the Heisenberg exchange coupling constants transform under (\ref{trans_orb}) as
\begin{equation}
\quad J_1^{\alpha,\beta}\rightarrow J_1^{p(\alpha),\beta}
\quad {\rm and} \quad J_2^{\alpha,\beta}\rightarrow J_2^{p(\alpha),p(\beta)}
\label{trans_exc}
\end{equation}
when orbital $\alpha$ lies on the $A (B)$ sublattice
of the N\'eel state.
Both $J_2^{\alpha,\beta}$ on the $B (A)$ sublattice 
and the Hund's rule coupling $J_0$ remain fixed.
Yet how does the spectrum of states transform under (\ref{trans_orb})?
Let $T_x$ and $T_y$ denote translations along the
$x$ axis and along the $y$ axis by one lattice constant $a$.  
Because $T_x T_y$
moves each sublattice to itself, 
it commutes with $P_A$ and with $P_B$.
If $\Psi$ has a crystal momentum $\hbar\bf k$, 
we therefore have that
$T_x T_y\, \Psi^{\prime} = e^{ i (k_x+ k_y) a}\, \Psi^{\prime}$.  
The crystal momentum then transforms under (\ref{trans_orb}) as 
\begin{equation}
{\bf k} \rightarrow {\bf k} \ {\rm or} \ {\bf k}  + (\pi / a, \pi / a) .
\label{trans_mmntm}
\end{equation}
Also, the spin of the new state $\Psi^{\prime}$ remains unchanged because $P_A$ and $P_B$ 
both commute
with ${\bf S}_i = \sum_{\alpha} {\bf S}_i (\alpha)$.

It is also useful to consider the special case where all nearest-neighbor and next-nearest-neighbor
exchange coupling constants are equal, respectively: 
$J_1^{\alpha,\beta} = J_1$ and $J_2^{\alpha,\beta} = J_2$.
The model Hamiltonian (\ref{j0j1j2}) then reduces to 
$H = {1\over 2} J_0 \sum_i {\bf S}_i \cdot {\bf S}_i +
    J_1\sum_{\langle i,j \rangle} {\bf S}_i \cdot {\bf S}_j +
    J_2\sum_{\langle\langle i,j \rangle\rangle} {\bf S}_i \cdot {\bf S}_j$.
Because ${\bf S}_i + {\bf S}_j$ commutes with ${\bf S}_i \cdot {\bf S}_i$,
the latter then commutes with the Hamiltonian.  
The total spin at a given site $i$ is hence a good quantum number.  
This means that the groundstate obeys Hund's rule in the classical limit
because states with maximum total spin at a given site 
minimize both the Hund's-rule energy ($J_0 < 0$) and the
Heisenberg exchange energies in such a case.
A violation of Hund's rule will therefore
require some variation in the  Heisenberg exchange coupling
constants among the different iron orbitals.

\section{Spin-1 Iron Atoms}
\begin{figure}
\includegraphics[scale=0.65, angle=-90]{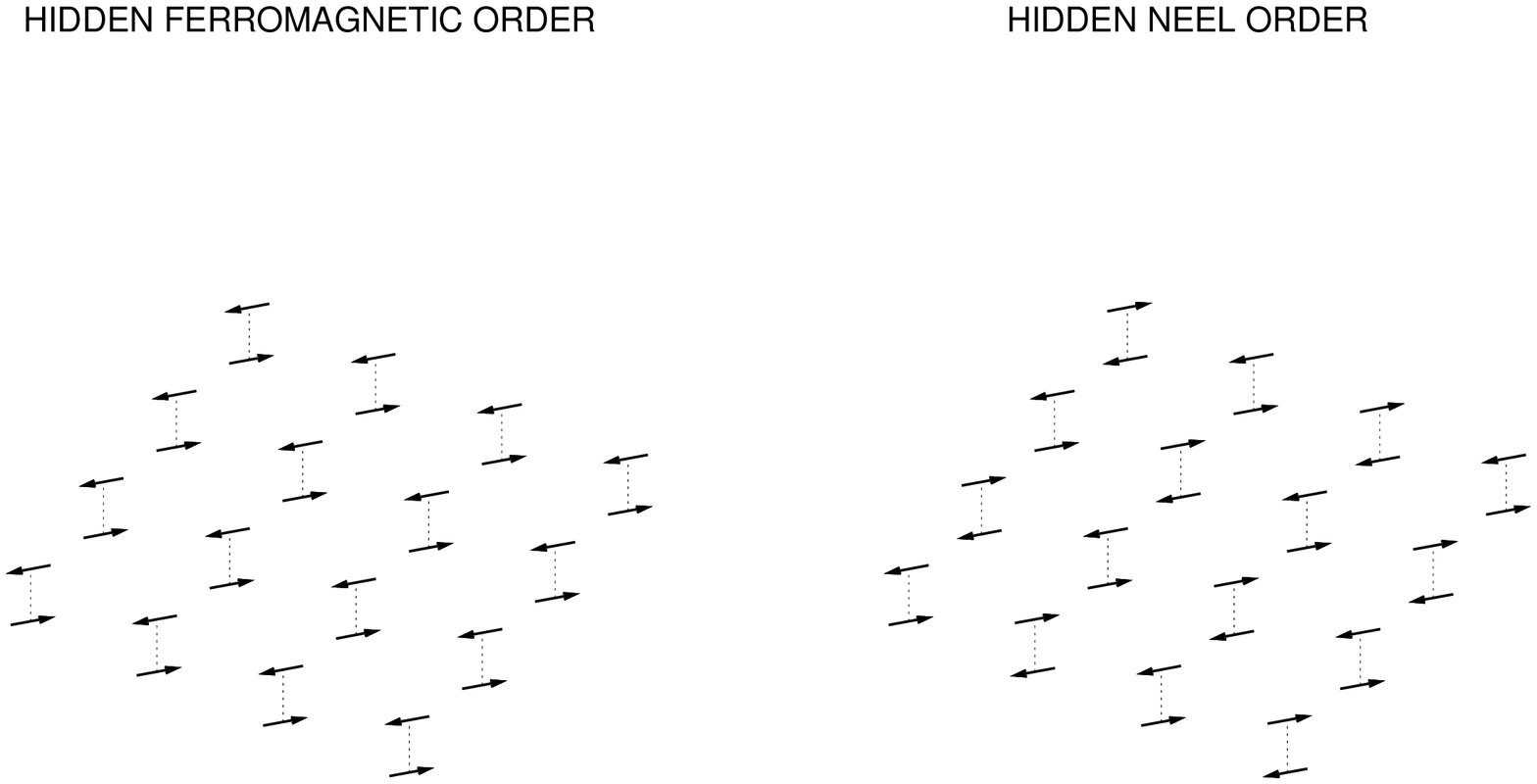}
\caption{Shown are two groundstates of a square lattice
of spin-1 iron atoms that violate Hund's rule, but that
exhibit hidden magnetic order. 
[See Eq. (\ref{j0j1j2}) and Table \ref{transitions}.]}
\label{hidden_order} 
\end{figure}

The iron 3d orbitals in ferro-pnictide materials experience a crystal field
due to the neigboring pnictide atoms that lie above/below the center of each unit
 square of iron atoms.
Both DFT calculations and basic considerations indicate that 
the crystal field splits these levels
so that the degenerate $3d_{xz}$ and $3d_{yz}$ orbitals 
lie just below the highest-energy $3d$ orbital.\cite{cao_08}$^,$\cite{li_08}
The Pauli principle then implies that 
spin $s = \hbar / 2$ moments exists
on both the $3d_{xz}$ and the $3d_{yz}$ orbitals
in the limit of weak Hund's rule coupling.
Henceforth,
we shall assume that only two spin-1/2 orbitals,
$a$ and $b$, 
exist per iron atom in parent compounds to ferro-pnictide high-$T_c$ superconductors.  
This is consistent with the ordered moment of $2\, \mu_B$ obtained by DFT calculations
for antiferromagnetic groundstates
of ferro-pnictide parent compounds.\cite{cao_08}$^,$\cite{dong_08}$^,$\cite{ma_08}
Let us next fix notation by assuming
unique diagonal and unique off-diagonal Heisenberg exchange coupling constants
over the nearest neighbor and the next-nearest neighbor links of
the square lattice of iron atoms:
%
\begin{equation}
J_{1(2)}^{a,a} = J_{1(2)}^{\parallel} = J_{1(2)}^{b,b}
\quad {\rm and} \quad
J_{1(2)}^{a,b} = J_{1(2)}^{\perp} = J_{1(2)}^{b,a}.
\label{jays}
\end{equation}
%
The author and Rezayi have recently shown that strong enough off-diagonal frustration,
$J_2^{\perp} > 0$,
can lead to either hidden ferromagnetic order or to hidden N\'eel order
among spin-1/2 moments that are governed by 
the $J_0$-$J_1$-$J_2$ model Hamiltonian\cite{jpr_ehr_09} (\ref{j0j1j2}).
Figure \ref{hidden_order} displays the two types of hidden order,
both of which show no net magnetic moment.
Proximity of the collinear antiferromagnetic state, 
which exists at $J_2 > |J_1|/2$ when Hund's rule is enforced,
to such hidden order can therefore
account for the low ordered moment that is observed by elastic neutron diffraction
in the cSDW state that exists at low temperature
in undoped ferro-pnictide materials.\cite{delacruz}
Balancing the energy of such hidden-order groundstates in the classical limit, $s\rightarrow\infty$,
against that of conventional magnetic order 
leads to critical values
for the Hund's rule coupling that are listed in Table \ref{transitions}.  
True magnetic order becomes unstable to hidden magnetic order
at weaker Hund's-rule coupling, $- J_0 < - J_{0c}$.
A true magnetically ordered groundstate therefore first becomes 
unstable to the hidden-order groundstate
that has the larger of the two critical values for the Hund's rule coupling strength, 
$- J_{0c}$.
By Table \ref{transitions}, for example, 
the true cSDW state becomes unstable to hidden ferromagnetic order at
weak Hund's rule coupling only if the off-diagonal nearest-neighbor
Heisenberg exchange is larger than the corresponding diagonal exchange:
$J_1^{\perp} > J_1^{\parallel}$.

\begin{table}
\begin{center}
\begin{tabular}{|c|c|c|c|}
\hline
                   & True cSDW & True N\'eel & True Ferromagnet \\
\hline
Hidden Ferromagnet & $- J_{0c} = 2 (J_1^{\perp} - J_1^{\parallel}) - 4 J_2^{\parallel}$
& $- J_{0c} = 4 J_2^{\perp} - 4 J_1^{\parallel}$ & $- J_{0c} = 4 J_2^{\perp} + 4 J_1^{\perp}$ \\
Hidden N\'eel & $- J_{0c} = 2 (J_1^{\parallel} - J_1^{\perp}) - 4 J_2^{\parallel}$
& $- J_{0c} = 4 J_2^{\perp} - 4 J_1^{\perp}$ & $- J_{0c} = 4 J_2^{\perp} + 4 J_1^{\parallel}$ \\
\hline
\end{tabular}
\caption{Listed are the phase boundaries that separate true magnetic order
from  hidden magnetic order.  The critical Hund's rule coupling
is obtained by balancing the classical energies of the model Hamiltonian (\ref{j0j1j2}).}
\label{transitions}
\end{center}
\end{table}

{\it Linear Spin-Wave Theory.}
Study of the classical limit, $s\rightarrow\infty$, therefore indicates that
extremely low ordered moments are possible at weak enough 
Hund's rule coupling, $J_0 < 0$,
when off-diagonal frustration exists:
$J_2^{\perp} > 0$.
The nature of spin-wave excitations about the hidden magnetic order 
that is shown in Fig. \ref{hidden_order}
can also be obtained from the classical limit.
In such case,
each spin-1/2 moment obeys the following dynamical equation for precession:
\begin{equation}
\dot {\bf S}_i (\alpha)  = 
\Biggl[{\partial H \over{\partial {\bf S}_i (\alpha)}} - {\bf h}_i (\alpha)\Biggr] \times {\bf S}_i (\alpha).
\label{precess}
\end{equation}
Here we have applied a magnetic field to the system, ${\bf h}_i (\alpha)$,
that is in principle sensitive to
the orbital degree of freedom, $\alpha = a$ or $b$.
If the external magnetic field is a plane-wave with wavenumbers ${\bf k}$ and frequency $\omega$,
then its transverse component with respect to the spin axis for hidden magnetic order
generates a corresponding linear response in the
transverse components of the electronic spin.
It has the form
\begin{equation}
{\bf S}_{\perp} (\pm, {\bf k}, \omega) = 
\chi_{\perp} (\pm, {\bf k},\omega) \, {\bf h}_{\perp} (\pm, {\bf k},\omega)
\label{diag_response}
\end{equation}
and
\begin{equation}
{\bf S}_{\perp} (\mp, {\bf k}, \omega) = 
i {\bar\chi}_{\perp} ({\bf k},\omega) \, {\bf n}_{\parallel} \times {\bf h}_{\perp} (\pm, {\bf k},\omega)
\label{off_diag_response}
\end{equation}
in the case of the hidden ferromagnetic state.
Here we define true ($+$) and hidden ($-$) magnetic moments and magnetic fields by
${\bf S} (\pm) = {\bf S}(a) \pm {\bf S}(b)$ and by
${\bf h} (\pm) = [{\bf h}(a) \pm {\bf h}(b)]/2$,
while ${\bf n}_{\parallel}$ denotes the  unit vector parallel to
the magnetization axis for hidden order.
Recall that the hidden ferromagnet is actually
an antiferromagnet over the two sublattices set by the two orbitals $a$ and $b$ 
(see Fig. \ref{hidden_order} and ref. \cite {af_sw_theory}).
The true magnetic susceptibility, $\chi_{\perp} (+, {\bf k},\omega)$,
the hidden magnetic susceptibility, $\chi_{\perp} (-, {\bf k},\omega)$,
and the conjugate susceptibility ${\bar\chi}_{\perp} ({\bf k},\omega)$
can then be obtained by solving the linear precession equation (\ref{precess}) 
that results in terms of the four transverse spin components per site,
${\bf S}_{\perp}(a)$ and ${\bf S}_{\perp}(b)$.
This yields
\begin{equation}
\chi_{\perp} (\pm, {\bf k},\omega) = s {\Omega_{\mp}\over{(\Omega_+ \Omega_-)^{1/2}}}
([\omega + (\Omega_+ \Omega_-)^{1/2}]^{-1} - [\omega - (\Omega_+ \Omega_-)^{1/2}]^{-1})
\label{chi}
\end{equation}
and
\begin{equation}
{\bar\chi}_{\perp} ({\bf k},\omega) = s {\omega\over{(\Omega_+ \Omega_-)^{1/2}}}
([\omega + (\Omega_+ \Omega_-)^{1/2}]^{-1} - [\omega - (\Omega_+ \Omega_-)^{1/2}]^{-1}),
\label{chi_bar}
\end{equation}
with frequencies given by
\begin{eqnarray*}
\Omega_- &=&  s(J_1^{\perp} - J_1^{\parallel}) \sum_{n=x,y} (2\, {\rm sin}\, {1\over 2} k_n a )^2 +
     s (J_2^{\perp} - J_2^{\parallel}) \sum_{n=+,-} (2\, {\rm sin}\, {1\over 2} k_n a)^2\\
\Omega_+ &=& 2sJ_0 + s J_1^{\perp} \sum_{n=x,y} (2\, {\rm cos}\, {1\over 2} k_n a )^2 
- s J_1^{\parallel} \sum_{n=x,y} (2\, {\rm sin}\, {1\over 2} k_n a )^2 \\
    && +  s J_2^{\perp} \sum_{n=+,-} (2\, {\rm cos}\, {1\over 2} k_n a)^2
  - s J_2^{\parallel} \sum_{n=+,-} (2\, {\rm sin}\, {1\over 2} k_n a)^2 
\end{eqnarray*}
in the case of the hidden ferromagnetic state.
Above, $k_\pm = k_x \pm k_y$,
$a$ denotes the square lattice constant,
and $s$ is the electron spin $\hbar / 2$.
Next observe that the hidden N\'eel state is related to the hidden ferromagnetic state
by the transformation (\ref{trans_orb}) mentioned in the previous section. 
(See Fig. \ref{hidden_order}.)
By (\ref{trans_exc}),
the frequencies $\Omega_{\pm}$ in that case are therefore obtained from the ones listed above
for the hidden ferromagnet after making the exchange
of the nearest-neighbor Heisenberg coupling constants,
$J_1^{\parallel} \leftrightarrow J_1^{\perp}$.
Further,
the wavenumbers of the hidden spin ${\bf S} (-)$ and 
of the hidden magnetic field ${\bf h} (-)$
that appear in the linear-response equations 
(\ref{diag_response}) and (\ref{off_diag_response})
must be replaced by the wave numbers
on the antiferromagnetic sublattice,
${\bf k} \rightarrow {\bf k} - (\pi / a, \pi / a)$.
The wavenumbers of the true spin and of  the true magnetic field that 
appear there
remain unchanged, however.
The end result is that the true magnetic susceptibility
$\chi_{\perp} (+, {\bf k}, \omega)$
in the hidden N\'eel state is given by that in the hidden ferromagnetic phase with
the  exchange
$J_1^{\parallel} \leftrightarrow J_1^{\perp}$ alone,
whereas the hidden magnetic susceptibility $\chi_{\perp} (-, {\bf k}, \omega)$ there
is given by that in the hidden ferromagnetic phase
with the previous exchange along with the replacement
${\bf k} \rightarrow {\bf k} + (\pi / a, \pi / a)$
in the wave numbers. [Cf. Eq. (\ref{trans_mmntm}).]

Equation (\ref{chi}) and the discussion above yield true static magnetic susceptibilities
$\chi_{\perp} (0) = (J_0 + 4 J_1^{\perp} + 4 J_2^{\perp})^{-1}$
for the hidden ferromagnetic state and
$\chi_{\perp} (0) = (J_0 + 4 J_1^{\parallel} + 4 J_2^{\perp})^{-1}$
for the hidden N\'eel state.
It also yields static susceptibilities  for hidden magnetic order given by
$\chi_{\perp} (-, {\bf k}, 0) = 2s / \Omega_-$,
which correctly diverge as $\bf k$ approaches the wavenumber that
corresponds to either long-range ferromagnetic or N\'eel order.
Finally, the poles of the dynamical susceptibility
for hidden magnetic order (\ref{chi}), $\chi_{\perp} (-, {\bf k}, \omega)$,
represent two spin-wave excitations of momentum $\hbar {\bf k}$ 
that correspond to the two possible orientations of the 
transverse component of the hidden moment,
${\bf S}_{\perp} (-, {\bf k}, \omega)$.
They both lie at an energy $\hbar (\Omega_+ \Omega_-)^{1/2}$ above the groundstate energy.
These spin-wave excitations are also reflected by 
the poles in the dynamical susceptibility for 
true magnetic order (\ref{chi}), $\chi_{\perp} (+, {\bf k}, \omega)$.
As mentioned above, however, the wave number ${\bf k}$ shifts 
by the characteristic amount $(\pi/a, \pi/a)$ 
in the case of hidden N\'eel order.
Last, both spinwaves have spectral weights
\begin{equation}
A ({\pm, \bf k}) = \pi s (\Omega_{\mp} / \Omega_{\pm})^{1/2}
\label{weight}
\end{equation}
in the true ($+$) and the hidden ($-$) channels
that are read off from the imaginary part of the dynamical susceptibility:
\begin{equation}
{\rm Im}\, \chi_{\perp} (\pm, {\bf k}, \omega) = 
A(\pm, {\bf k}) (\delta[\omega - (\Omega_+ \Omega_-)^{1/2}] 
- \delta[\omega + (\Omega_+ \Omega_-)^{1/2}]).
\label{im_chi}
\end{equation}
The peaks in the dynamical structure function above
signal spin-wave excitations that
are observable by inelastic neutron scattering
only in the true magnetic ($+$) channel.  
Recalling the discussion following
Eq. (\ref{chi}), the true dynamical susceptibility $\chi_{\perp} (+, {\bf k}, \omega)$
is common to a hidden ferromagnet and to a hidden N\'eel state that are related by
the transformation (\ref{trans_orb}).
Inelastic neutron scattering is therefore unable to discriminate between these two possible
types of hidden magnetic order!

\begin{figure}
\includegraphics[scale=0.65, angle=-90]{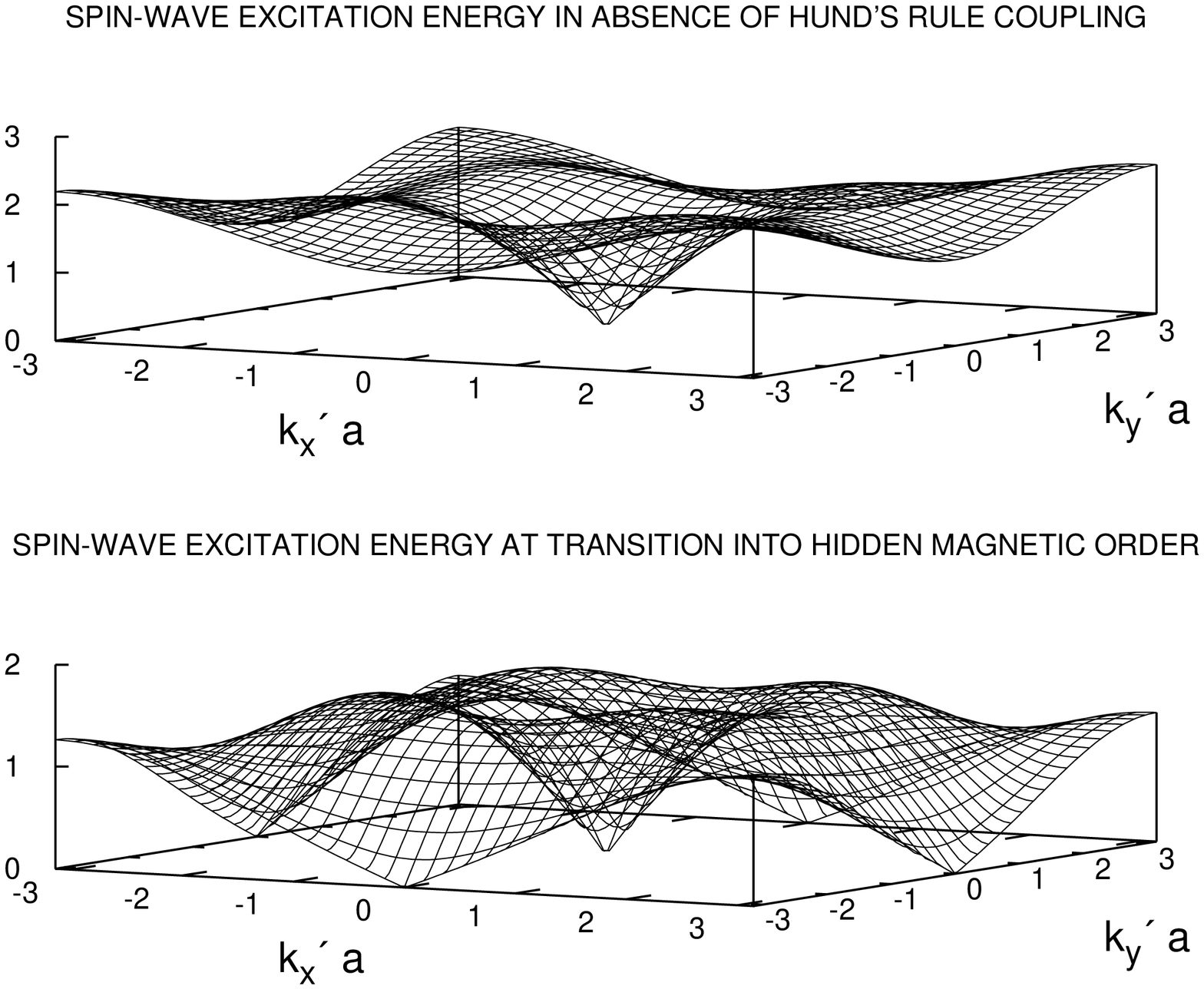}
\caption{Displayed is the linear spin-wave spectrum of the Hamiltonian (\ref{j0j1j2})
for the following set of Heisenberg exchange coupling constants:
$J_1^{\parallel (\perp )} = 0$, $J_1^{\perp (\parallel)} > 0$, 
and $J_2^{\parallel} = 0.3\, J_1^{\perp (\parallel )} = J_2^{\perp}$.
The Hund's rule coupling at the quantum critical point that separates 
hidden ferromagnetic (N\'eel) order
from a cSDW is set by the value listed in Table \ref{transitions}.
Spin-wave energies are given in units of $\hbar^2 J_1^{\perp (\parallel )}$
and the wavenumber ${\bf k}^{\prime}$ is associated with the antiferromagnetic sublattice
for hidden order. (See Fig. \ref{hidden_order}.)}
\label{sw}
\end{figure}

\begin{figure}
\includegraphics[scale=0.65, angle=-90]{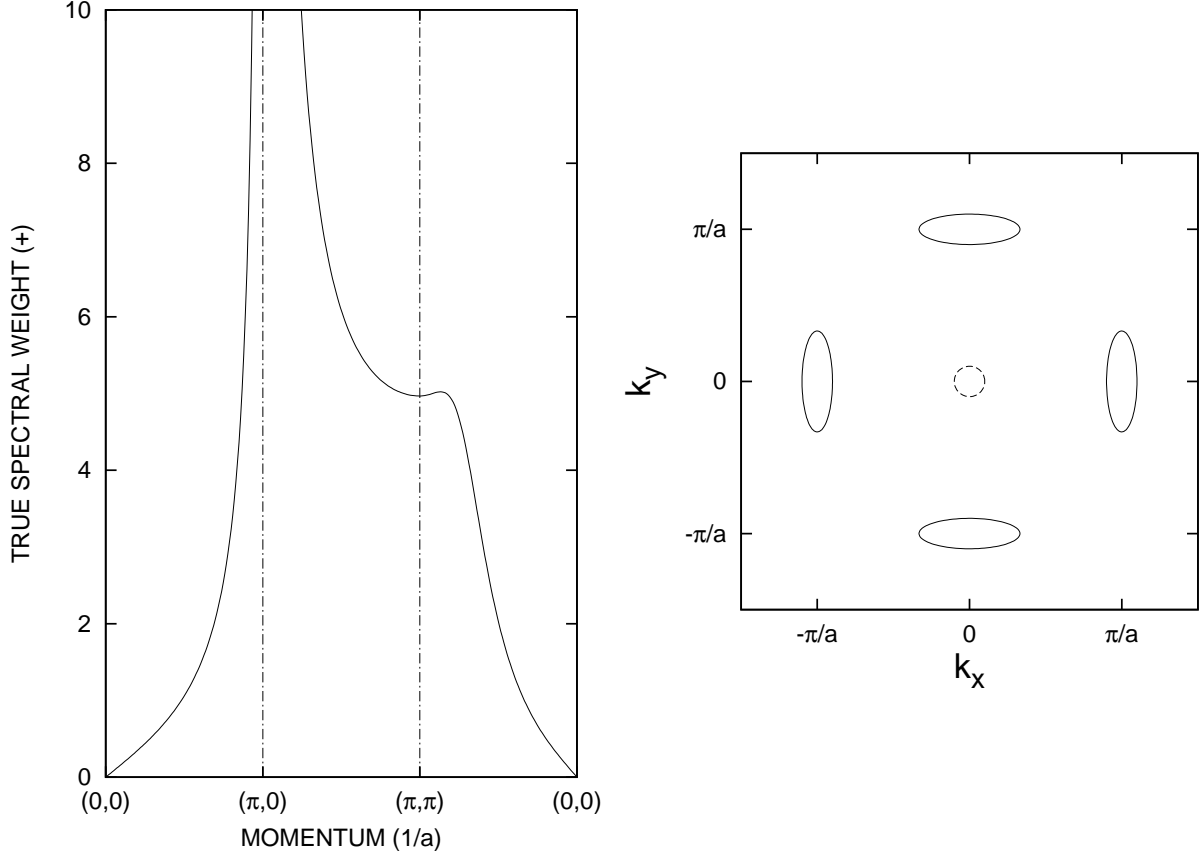}
\caption{The spectral weight of observable (+) spinwaves at
the quantum critical point that separates hidden ferromagnetic (N\'eel) order from a cSDW
is shown in units of $\hbar$.
The following set of Heisenberg exchange coupling constants are used:
$J_1^{\parallel (\perp )} = 0$, $J_1^{\perp (\parallel )} > 0$,
and $J_2^{\parallel} = 0.3\, J_1^{\perp (\parallel )} = J_2^{\perp}$.
The critical Hund's rule exchange coupling is set to the value listed in Table \ref{transitions}.
Also shown are contours over the Brillouin zone for such spinwaves at a low fixed excitation energy.}
\label{cp_spctrm}
\end{figure}

Figure \ref{sw} depicts the linear spin-wave spectrum for
a cSDW with intervening  hidden ferromagnetic (N\'eel)  order
at two extreme values for the Hund's rule coupling: absent and critical.  
Here, we have chosen Heisenberg exchange coupling constants
$J_1^{\parallel (\perp)} = 0, J_1^{\perp (\parallel)} > 0$, 
and $J_2^{\parallel}  = 0.3\, J_1^{\perp (\parallel)} = J_2^{\perp}$,
which lies near maximum frustration when Hund's rule is obeyed.
The top graph displays how the spectrum collapses linearly to the groundstate energy
at zero momentum in the absence of Hund's rule coupling.
In general,
the spin-wave velocity is given by
\begin{equation}
v_0 = 2 s a ([J_1^{\perp} - J_1^{\parallel} + 2 (J_2^{\perp} - J_2^{\parallel})]
\cdot [{1\over 2} J_0 + 2 J_1^{\perp} + 2 J_2^{\perp}])^{1/2}
\label{sw_vel}
\end{equation}
in the hidden ferromagnet, 
and by the same expression after the exchange (\ref{trans_orb})
$J_1^{\parallel} \leftrightarrow J_1^{\perp}$
in the hidden N\'eel state.
It correctly collapses to zero at the phase boundary with the true ferromagnetic state
by Table \ref{transitions},
at which point the spin-wave excitation frequency disperses quadratically with wavenumber.
Notice, however, the mild depression in the top spectrum
that is  centered at the cSDW wave number ${\bf k}_{cSDW} = (\pi/a, 0)$.
The bottom graph in Fig. \ref{sw} shows how it softens completely to zero energy
at the critical value for Hund's rule coupling, $- J_{0c} = 0.8\, J_1^{\perp (\parallel )}$,
in which case the hidden ferromagnetic (N\'eel) state
transits into the cSDW state by Table \ref{transitions}. 
These special zero-energy
modes result from the collapse of the frequency
$\Omega_+ ({\bf k}_{cSDW})$ to zero there.
Evaluating $(\Omega_+ \Omega_-)^{1/2}$ at ${\bf k}_{cSDW}$ from  the expressions for the frequencies
that are listed below Eq. (\ref{chi_bar}) yields the spin gap there:
\begin{equation}
\Delta_{cSDW} = 2s [(4 J_2^{\perp} - J_{0c}) (J_0 - J_{0c})]^{1/2}.
\label{sdw_gap}
\end{equation}
The former also implies a divergent
spectral weight (\ref{weight})
for such  spinwaves at the quantum phase transition:
\begin{equation}
A(+, {\bf k}_{cSDW}) = \pi s [(4 J_2^{\perp} - J_{0c}) / (J_0 - J_{0c})]^{1/2}.
\label{weight_sdw}
\end{equation}
Because inelastic neutron scattering is proportional to
${\rm Im}\, \chi_{\perp} (+, {\bf k}, \omega)$,
low-energy spin-waves with momentum
near $\hbar {\bf k}_{cSDW}$ will dominate such measurements
near the transition between 
the cSDW state and hidden magnetic order (see Fig. \ref{cp_spctrm}). 
Study of the expressions for the frequencies $\Omega_{\pm}$
that are listed below Eq. (\ref{chi_bar})
yields that 
the dominant spin-wave excitations centered at the cSDW wave numbers ${\bf k}_{cSDW}$
disperse up from the groundstate energy as
$[v_{l}^2 (k_{l} - \pi / a)^2 + v_{t}^2 k_{t}^2 + \Delta_{cSDW}^2]^{1/2}$
in general.  (See Fig. \ref{sw}.)
Here $k_l$ and $k_t$ are the components of ${\bf k}$ that are respectively 
parallel and perpendicular to ${\bf k}_{cSDW}$. 
Near criticality, $\Delta_{cSDW} \rightarrow 0$,
the cSDW order is short range, with correlations lengths
$\xi_{cSDW}^{(l)} =  v_l / \Delta_{cSDW}$ and
$\xi_{cSDW}^{(t)} =  v_t / \Delta_{cSDW}$
in the longitudinal and transverse directions.
At criticality,
$\Delta_{cSDW} = 0$,
the longitudinal velocity $v_{l}$ coincides with the spin-wave velocity $v_0$ 
for hidden magnetic order  (\ref{sw_vel}),
while the  transverse velocity $v_{t}$ is
equal to the previous divided by the anisotropy parameter
\begin{equation}
{v_{l} \over{v_{t}}} = 
\Biggl[{2(J_2^{\parallel} + J_2^{\perp}) + J_1^{\parallel} + J_1^{\perp}
\over{2(J_2^{\parallel} + J_2^{\perp}) - J_1^{\parallel} - J_1^{\perp}}}\Biggr]^{1/2}.
\label{anisotropy}
\end{equation}
The latter coincides with the anisotropy of low-energy spinwaves that is 
predicted by the conventional $J_1$-$J_2$ model, in which case 
$J_{1(2)} = [J_{1(2)}^{\parallel} + J_{1(2)}^{\perp}]/2$
by Hund's rule.

{\it Exact Diagonalization.}
The  above semi-classical analysis indicates that large enough  off-diagonal frustration 
can induce a  quantum phase transition into hidden magnetic order that is unfrustrated, 
but that violates Hund's rule.  
We shall now confirm this by obtaining
the exact low-energy spectrum of states for the 
$J_0$-$J_1$-$J_2$ model  (\ref{j0j1j2})
with two spin-1/2 moments per site
over a 4 by 4 square lattice with periodic boundary conditions.
The model Hamiltonian operator (\ref{j0j1j2})  acts on
a Hilbert space that is restricted to have 16 up spins and 16 down spins.
Next, translational invariance is exploited in order to 
bring $H$ into block-diagonal form.
Each block of the Hamiltonian is labeled by the allowed momentum quantum numbers: 
$(k_x a, k_y a) =
(0,0)$, $(\pi,0)$, $(\pi,\pi)$, $(\pi / 2, 0)$, $(\pi / 2, \pi / 2)$ and $(\pi, \pi / 2)$, 
plus their symmetric counterparts.  
Spin-flip symmetry 
is also exploited to further block-diagonalize the Hamiltonian at 
such momenta into two blocks
that are respectively even and odd under it.
The combination of translation and spin-flip symmetries reduces the dimension of each
block  to a little under 19,000,000 states.
Next, we construct the Hamiltonian operator (\ref{j0j1j2}) over each of these subspaces.
Every term in Hamiltonian (\ref{j0j1j2}) represents a
Heisenberg spin exchange that permutes the Bloch-wave type states.
These permutations are stored in memory. 
Hamiltonian matrix elements that involve Bloch waves 
composed of configurations of spin up and spin down
that display absolutely no non-trivial translation invariance 
up to a spin-flip
are also stored in memory. 
They are calculated otherwise.  
Because the vast majority of Bloch waves lie in the first category,
the Hamiltonian operator is stored in memory for all practical purposes.
This speeds up its application on a given state tremendously.
The application of the Hamiltonian $H$ on a given state is
accelerated further by
enabling  shared-memory parallel computation through OpenMP directives.
Last, we apply the iterative  Lanczos technique numerically over each subspace
labeled by momentum and by spin-flip quantum numbers.\cite{lanczos}
The ARPACK subroutine library is employed for this purpose.\cite{arpack}

\begin{figure}
\includegraphics[scale=0.65, angle=-90]{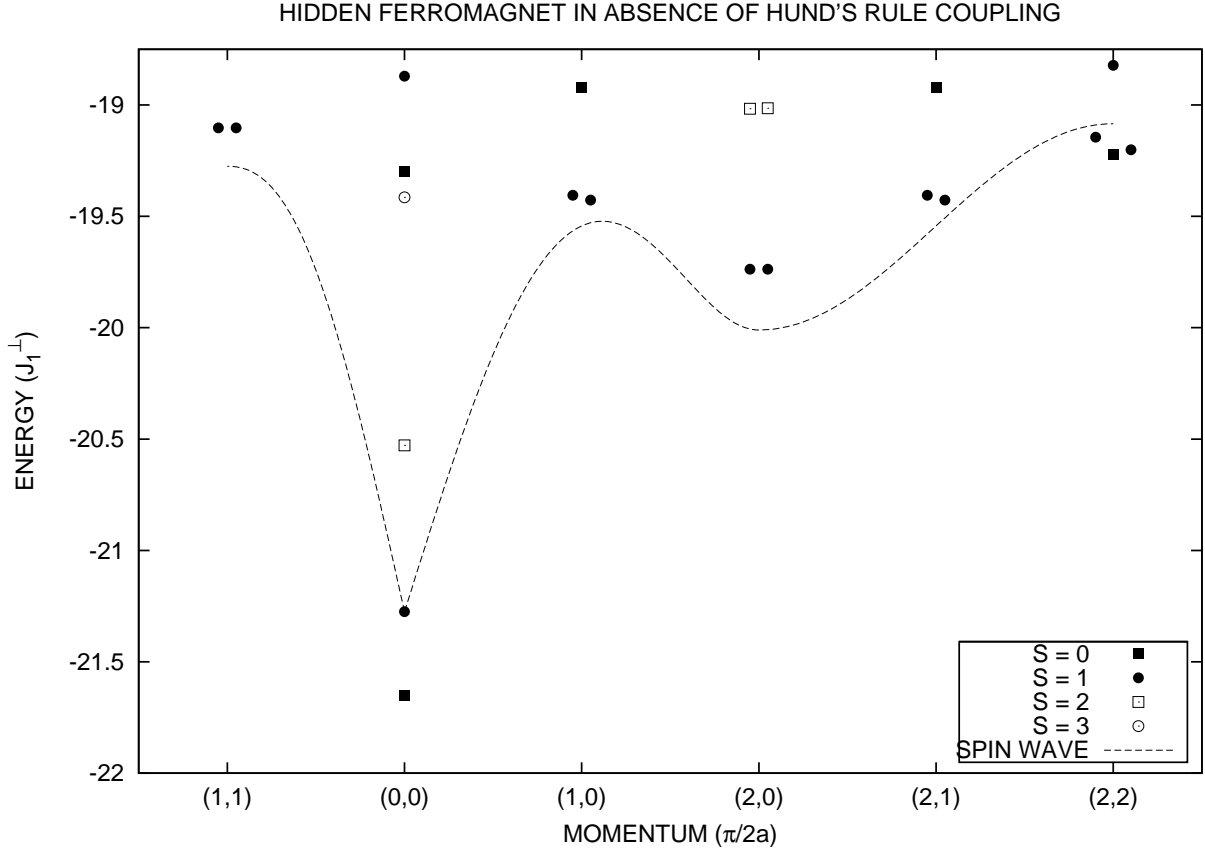}
\caption{Shown is the low-energy spectrum for $4\times 4\times 2$ spin-1/2 moments
that experience the following set of Heisenberg exchange coupling constants:
$J_0 = 0$, $J_1^{\parallel} = 0$, $J_1^{\perp} > 0$,
and $J_2^{\parallel} = 0.3\, J_1^{\perp} = J_2^{\perp}$.
The lowest-energy spin-1 state at zero  momentum is used as the reference
for the linear spin-wave approximation.
Hereafter, we set $\hbar \rightarrow 1$.}
\label{0_hund_spectra}
\end{figure}

\begin{figure}
\includegraphics[scale=0.65, angle=-90]{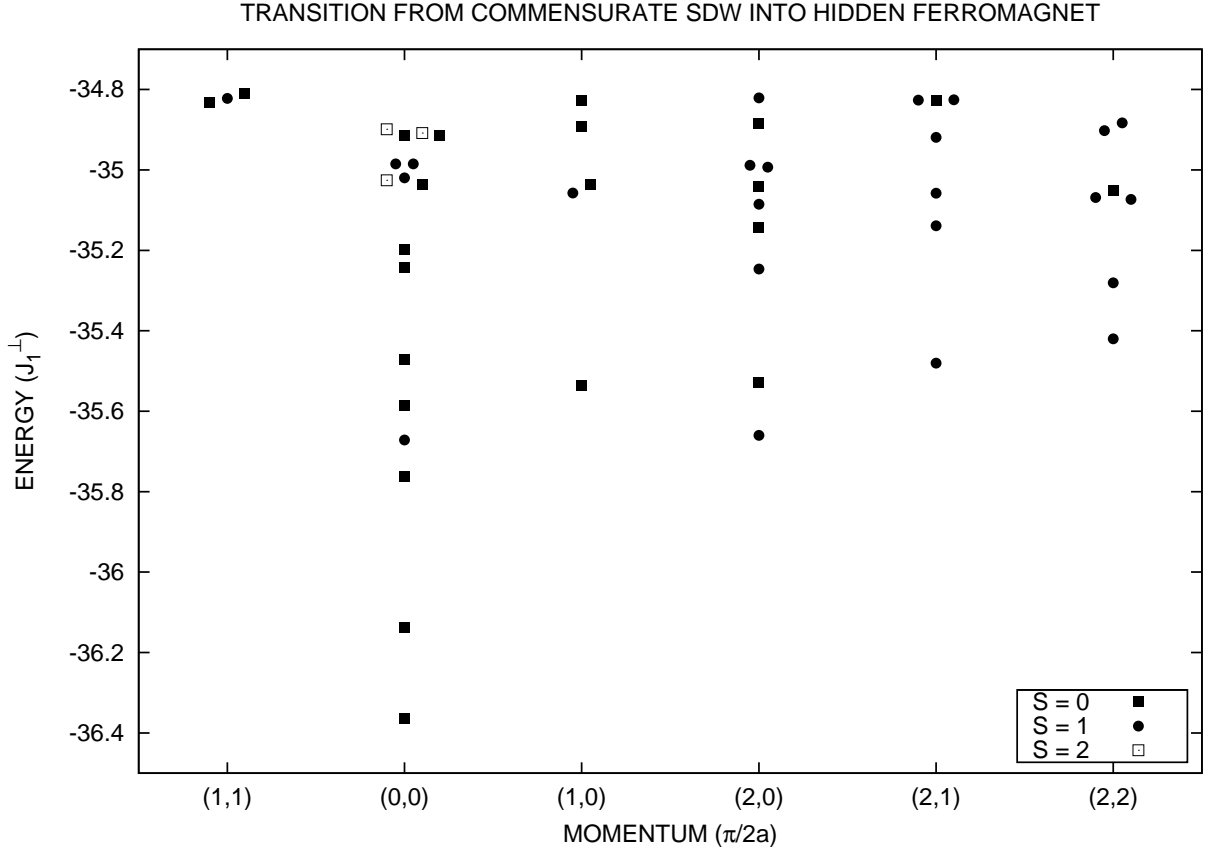}
\caption{Shown is the low-energy spectrum for 
$4\times 4\times 2$ spin-1/2 moments
that experience the following set of Heisenberg exchange coupling constants:
$J_1^{\parallel} = 0$, $J_1^{\perp} > 0$,
and $J_2^{\parallel} = 0.3\, J_1^{\perp} = J_2^{\perp}$.
The Hund's rule exchange coupling is set to $J_0 = - 1.35\, J_1^{\perp}$.}
\label{cp_hund_spectra}
\end{figure}

\begin{figure}
\includegraphics[scale=0.65, angle=-90]{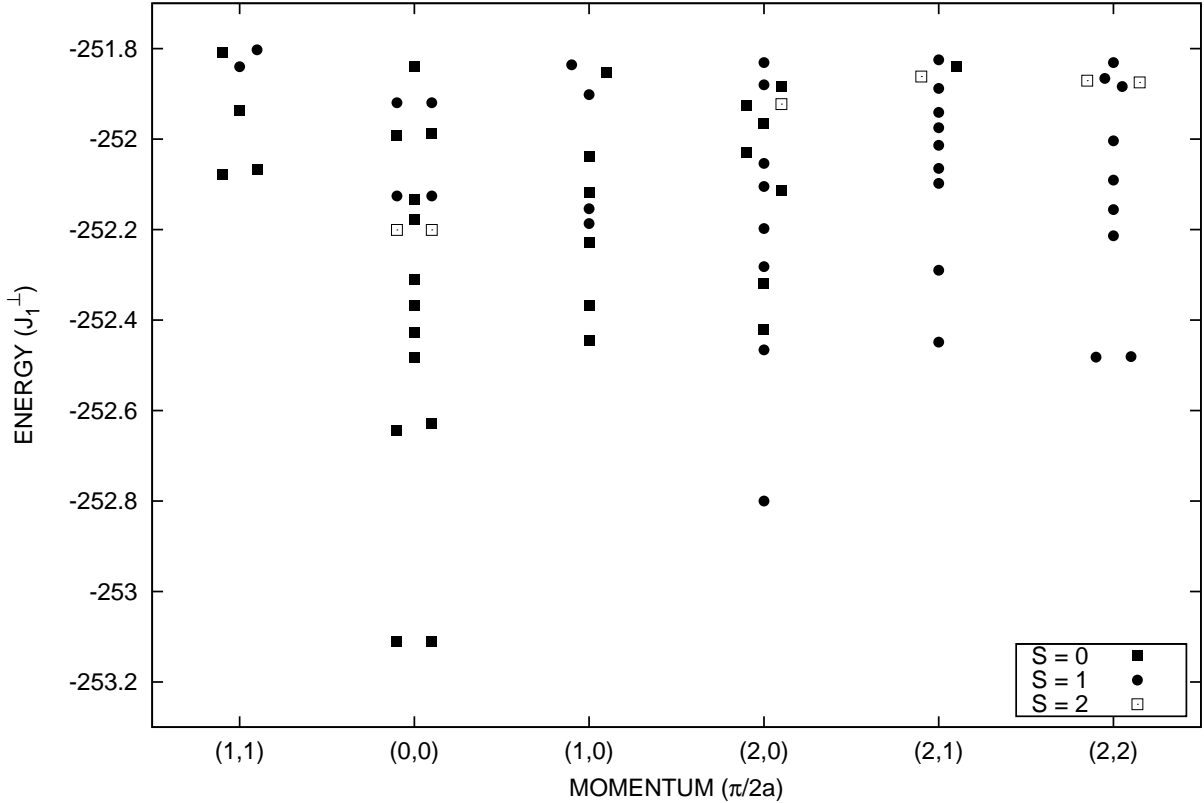}
\caption{The low-energy spectrum of the cSDW is obtained from exact diagonalization of 
$4\times 4\times 2$ spin-1/2 moments.  The following set of Heisenberg exchange coupling
constants are  used: $J_1^{\parallel} = 0$, $J_1^{\perp} > 0$,
$J_2^{\parallel} = 0.3\, J_1^{\perp} = J_2^{\perp}$,
and $J_0 = - 15\, J_1^{\perp}$.}
\label{sdw_spectra}
\end{figure}

\begin{figure}
\includegraphics[scale=0.65, angle=-90]{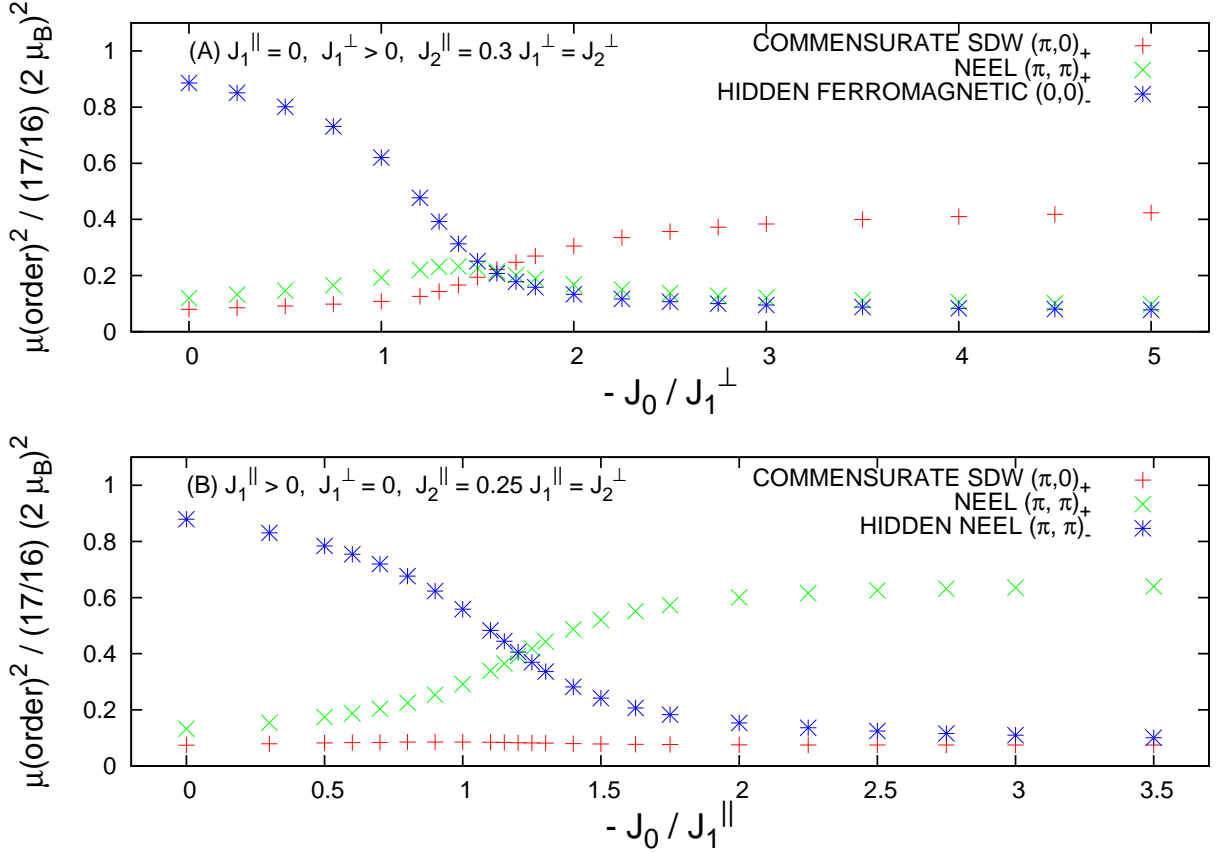}
\caption{Displayed is the autocorrelation of the order parameter (\ref{OP})
for true (+) and for hidden (-) magnetic order 
on a $4 \times 4 \times 2$ lattice of spin-1/2 moments 
as a function of Hund's rule coupling.
It is normalized to its value in the true ferromagnetic state (see text).}
\label{order_moment_1}
\end{figure}

Figures \ref{0_hund_spectra}, \ref{cp_hund_spectra} and \ref{sdw_spectra}
show how the low-energy spectrum of the $J_0$-$J_1$-$J_2$ model  (\ref{j0j1j2})
evolves with the strength of the Hund's rule coupling 
in the presence of off-diagonal frustration.
The following Heisenberg exchange coupling constants were chosen:
$J_1^{\parallel} = 0$, $J_1^{\perp} > 0$, and $J_2^{\parallel} = 0.3\, J_1^{\perp} = J_2^{\perp}$.
A hidden ferromagnetic state is then expected in the absence of Hund's rule coupling,
while a cSDW is expected when Hund's rule in obeyed.
In the former case displayed by Fig. \ref{0_hund_spectra},
notice the coincidence
between the  linear spin-wave  approximation 
about hidden-order,  Eq. (\ref{im_chi}) and Fig. \ref{sw},
and the present exact-diagonalization results.
It indicates that the classical prediction of long-range hidden magnetic order
holds true for spin-1/2 moments per iron orbital. 
Figure \ref{sdw_spectra} displays the spectrum
at strong Hund's rule coupling,  $ - J_0 = 15\, J_1^{\perp}$, on the other hand.
The bulk of the groundstate energy is due to  the Hund's rule coupling, 
which is given by
$16\, J_0 = - 240\, J_1^{\perp}$ when Hund's rule is obeyed.
Notice the low-energy spin-1 state at the cSDW wave number $(\pi/a, 0)$,
as well as the softening of the  spin-1 excited state
at the wave number $(\pi/a, \pi/a)$ that characterizes  the N\'eel state.
The dispersion of these spin-1 excited states resembles
the spin-wave spectrum predicted for the cSDW in the conventional $J_1$-$J_2$ model,
where Hund's rule is obeyed.\cite{oitmaa_10}$^,$\cite{zhao_09}
Figure \ref{sdw_spectra} also displays that the groundstate is doubly degenerate, 
which again is consistent with cSDW order over the square lattice.
Last, Fig. \ref{cp_hund_spectra} shows
the low-energy spectrum
for Hund's rule coupling $J_0 = - 1.35\, J_1^{\perp}$
near a possible  quantum phase transition (cf. Fig. \ref{order_moment_1}).
The lowest-energy spin-1 excitations at momentum
$(0, 0)$ and $(\pi/a, 0)$ are degenerate, and their dispersion resembles the
prediction from linear spin-wave theory that is shown by Fig. \ref{sw}.
Notably,
the lowest-energy spin-1 excitation
does {\it not} ``dip down'' at momentum $(\pi/a, \pi/a)$.
Table \ref{transitions} predicts a transition into hidden order at
$J_0 = - 0.8\, J_1^{\perp}$ in the classical limit at large spin $s$, however,
which suggests that quantum effects renormalize up the critical Hund's rule coupling
in the physically relevant case of spin-1/2 iron $3d$ orbitals.

Figure  \ref{order_moment_1}
shows the evolution of relevant magnetic order parameters with Hund's rule coupling
under the previous set of Heisenberg exchange coupling constants.
Plotted on the vertical axis is the autocorrelation
$\langle {\bf O} ({\bf k})_\pm \cdot {\bf O} (-{\bf k})_\pm \rangle_0$
of the order parameter
\begin{equation}
{\bf O} ({\bf k})_\pm = \sum_i e^{i{\bf k}\cdot{\bf r}_i} [{\bf S}_i(a) \pm {\bf S}_i(b)]
\label{OP}
\end{equation}
over the groundstate normalized to its value in the true ferromagnetic state,
$\langle {\bf O} (0)_+ \cdot {\bf O} (0)_+ \rangle_0 = 16\cdot 17\,\hbar^2$. 
The relation to the ordered magnetic moment per iron atom, 
$\mbox{\boldmath$\mu$} (\pm, {\bf k})$, 
given there is obtained from the identity
$| 16\, \mbox{\boldmath$\mu$} (\pm, {\bf k})|^2 
 = 
(2\mu_B/\hbar)^2 \langle {\bf O} ({\bf k})_\pm \cdot {\bf O} (-{\bf k})_\pm \rangle_0$
that is valid over the 4 by 4 square lattice of spin-1 iron atoms.
The top graph in Fig.   \ref{order_moment_1}
displays how the square of the ordered moment for true cSDW order (+)
decays with decreasing Hund's rule coupling.
It   notably accounts for the low ordered moment that is observed
by elastic neutron diffraction in an undoped parent compound to the 
recently discovered ferro-pnictide
high-$T_c$ superconductors.\cite{delacruz}
Also shown there is how hidden ferromagnetic order is established once cSDW order fades away.
Figure  \ref{order_moment_1}
then provides evidence for a quantum phase transition that seperates a hidden ferromagnetic state
at weak Hund's rule coupling from a cSDW state at strong Hund's rule coupling.
Notice that the putative quantum phase transition at $J_0 \cong -1.35\, J_1^{\perp}$
(see Fig. \ref{cp_hund_spectra})
occurs at a Hund's rule exchange coupling that is almost a factor-of-2 times larger
than the classical prediction
$J_{0c} = - 0.8\, J_1^{\perp}$ listed in Table \ref{transitions}.
Last, the bottom graph in Fig.  \ref{order_moment_1} shows the evolution of the relevant
order parameters in the case of true N\'eel order with intervening hidden N\'eel order.
Here, we have chosen Heisenberg exchange coupling constants
$J_1^{\parallel} > 0, J_1^{\perp} = 0$, 
and $J_2^{\parallel}  = 0.25\, J_1^{\parallel} = J_2^{\perp}$,
which lies at maximum frustration when Hund's rule is obeyed.
The common inflection point at $J_0 \cong - 1.2 J_1^{\parallel}$
is consistent with a quantum phase transition at a value of Hund's rule
coupling that is noticably larger than the  classical prediction of $J_{0c} = - J_1^{\parallel}$ 
listed in Table \ref{transitions}.

\section {Discussion and Conclusions}
\begin{figure}
\includegraphics[scale=0.65, angle=-90]{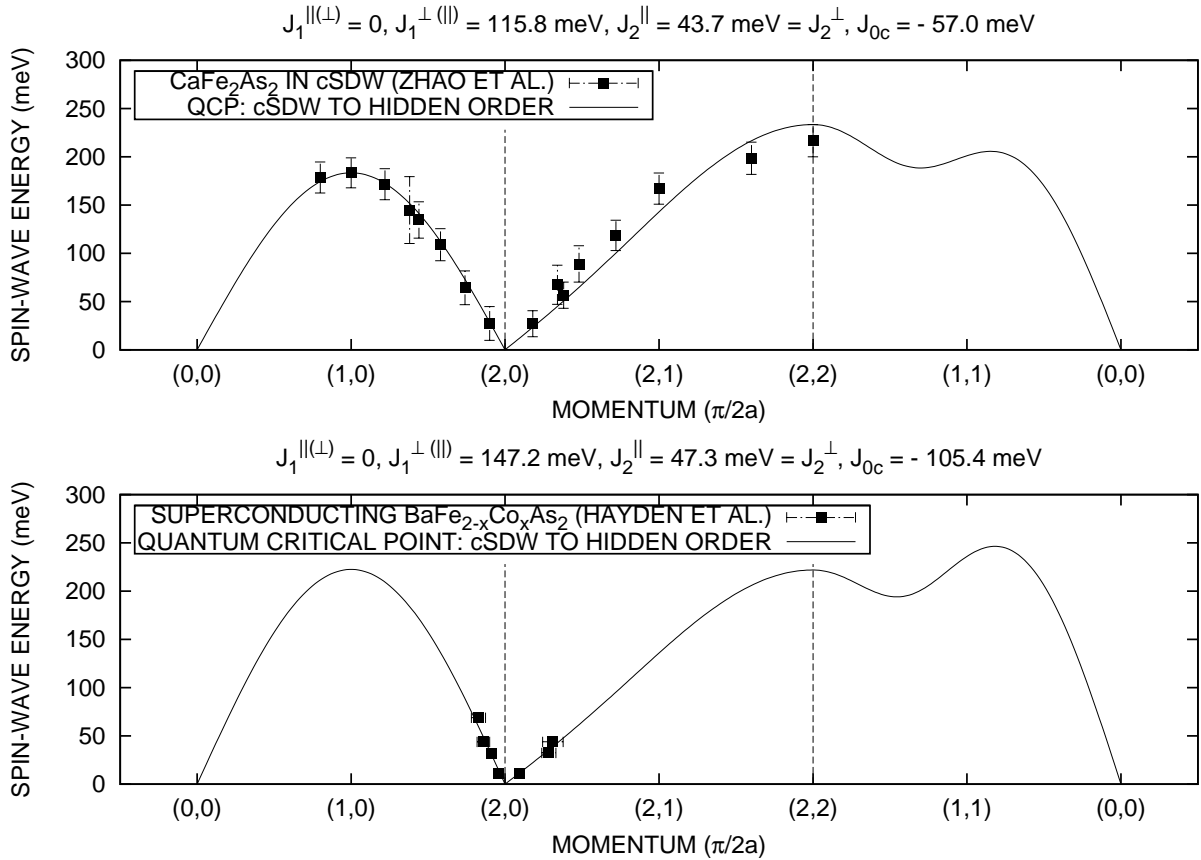}
\caption{The spin excitation spectra of two pnictide materials are fit to the linear spinwave
spectrum expected at a quantum critical point (QCP) that seperates a cSDW from ferromagnetic 
(N\'eel) hidden order. [See Eq. (\ref{im_chi}) and Fig. \ref{sw}.] 
The wavenumber in the direction perpendicular to the planes of iron atoms
corresponds to antiferromagnetic order
in  both sets of data.
(See refs. \cite{zhao_09} and \cite{hayden_10}.)}
\label{qcp_fit}
\end{figure}

Recent inelastic neutron scattering measurements 
on the parent compound to ferro-pnictide
superconductors CaFe$_2$As$_2$
have uncovered well-defined spin-wave excitations 
in the cSDW phase
throughout virtually the entire Brillouin zone.\cite{zhao_09}
Instead of softening 
at the wavenumber $(\pi/a, \pi/a)$ that corresponds to N\'eel order
as predicted by the conventional $J_1$-$J_2$ model over the square lattice,
they show a local maximum there.
Figure \ref{qcp_fit} displays a fit of the spin-wave spectrum obtained by inelastic neutron
scattering on CaFe$_2$As$_2$ to the prediction for the linear spin-wave dispersion of a
square lattice of spin-1 iron atoms, $(\Omega_+ \Omega_-)^{1/2}$,
at the quantum critical point that seperates
the cSDW phase from hidden ferromagnetic (N\'eel) order 
[see Eq. (\ref{im_chi}) and Fig. \ref{sw}].  
The spin per orbital is set to $s = 1/2$ and
the following Heisenberg exchange coupling
constants are used: $J_{0c} = - 57.0$ meV, $J_1^{\parallel (\perp)} = 0$,
$J_1^{\perp (\parallel)} = 115.8$ meV,
and $J_2^{\parallel} = 43.7\, {\rm meV}\, = J_2^{\perp}$.
These imply a ratio $J_2 / J_1 = 0.75$ between the next-nearest-neighbor and 
the nearest-neighbor Heisenberg exchange constants deep inside the cSDW,
where Hund's rule is obeyed.
Notice that the measured spin-wave spectrum terminates before it reaches the predicted
Goldstone modes at zero momentum that correspond to hidden magnetic order.
This is consistent with Fig. \ref{cp_spctrm},
which displays how the spectral weight of low-energy spin-waves about zero momentum
is dwarfed by that (\ref{weight_sdw}) of low-energy spinwaves about momentum $(\pi/a,0)$
at the quantum critical point that seperates the cSDW phase from hidden magnetic order.

More recent inelastic neutron scattering measurements on the ferro-pnictide superconductor
BaFe$_{2-x}$Co$_x$As$_2$ have uncovered low-energy  spin-wave excitations
that disperse {\it up} in energy in an anisotropic fashion
away from the wavenumbers that correspond to cSDW order;\cite{hayden_10} e.g. $(\pi/a, 0)$.
Paradoxically, no evidence for long-range magnetic order exists in this ferro-pnictide superconductor.
Figure \ref{qcp_fit} displays a fit of the spin-wave spectrum obtained by inelastic neutron
scattering on BaFe$_{2-x}$Co$_x$As$_2$ to the prediction for the linear spin-wave dispersion of a
square lattice of spin-1 iron atoms, $(\Omega_+ \Omega_-)^{1/2}$,
at the quantum critical point that seperates
the cSDW phase from hidden ferromagnetic (N\'eel) order 
[see Eq. (\ref{im_chi}) and Fig. \ref{sw}].
The spin per orbital is again set to $s = 1/2$,
and the following Heisenberg exchange coupling
constants are used: $J_{0c} = - 105.4$ meV, $J_1^{\parallel (\perp)} = 0$,
$J_1^{\perp (\parallel)} = 147.2$ meV,
and $J_2^{\parallel} = 47.3\, {\rm meV}\, = J_2^{\perp}$.
This fit implies a ratio $J_2 / J_1 = 0.64$ between the next-nearest-neighbor and 
the nearest-neighbor Heisenberg exchange constants deep inside the cSDW.
Although the conventional $J_1$-$J_2$ model over the square lattice
can also account for the linear and anisotropic
dispersion of the spinwaves that are
observed in superconducting\cite{hayden_10} BaFe$_{2-x}$Co$_x$As$_2$,
it cannot account for the absence of long-range magnetic order.
Both the semi-classical analysis about hidden magnetic order and
the exact diagonalization studies on a $4 \times 4$ lattice
described in the previous section
indicate, on the other hand,
that the moment associated with true cSDW order is extremely small
at the quantum critical point into hidden magnetic order.
(See Figs. \ref{hidden_order} and \ref{order_moment_1}.)

In conclusion, 
a linear spin-wave analysis of the $J_0$-$J_1$-$J_2$ model Hamiltonian (\ref{j0j1j2})
that describes a square lattice of frustrated spin-1 iron atoms 
yields a quantum critical point as a function of Hund's rule exchange coupling ($- J_0$) 
that seperates a cSDW from hidden ferromagnetic or hidden N\'eel order.
The observable spectral weight of the Goldstone modes associated with hidden order
is predicted to be small in general
(see  Fig. \ref{cp_spctrm}). 
A strong variation between the diagonal and the off-diagonal 
exchange couplings of the multi-orbital Heisenberg model (\ref{j0j1j2})
across nearest-neighbors {\it alone} is  sufficient
for Hund's rule to give way to hidden magnetic order.
Recent DFT calculations obtain ferromagnetic  direct exchange and 
antiferromagnetic superexchange across nearest-neighbor iron moments in 
the cSDW phase of ferro-pnictide materials.\cite{ma_08}
Their superposition potentially can result in 
the required difference between $J_1^{\parallel}$ and $J_1^{\perp}$.
Fits of the critical linear spin-wave spectrum 
to recent experimental studies of magnetic excitations
in ferro-pnictide materials\cite{zhao_09}$^,$\cite{hayden_10}
yield a Hund's rule exchange coupling constant on the order of $100$ meV.
Figure \ref{order_moment_1} displays 
the dependence of the ordered magnetic  moments on Hund's rule coupling
obtained from exact diagonalization of the model Hamiltonian (\ref{j0j1j2})
over a $4 \times 4$ lattice, with two spin-1/2 orbitals per site.
The top graph implies a critical Hund's rule coupling 
at the transition between hidden ferromagnetic order and  cSDW order
that is almost twice the
classical prediction listed in Table \ref{transitions}:
$- J_{0c} = 0.8\, J_1^{\perp}$. 
Correcting for this discrepancy brings the value 
of the Hund's rule coupling closer to the estimate
of 700 meV that is obtained for ferro-pnictides 
by dynamical mean field theory.\cite{haule_09}
(Note that $-J_0$ is twice as big as the Hund's rule exchange coupling that 
appears naturally in multi-orbital Hubbard models.)

The success of the theoretical fits to the inelastic neutron scattering data 
displayed by Fig. \ref{qcp_fit}
indicates that a local-moment picture for magnetism in ferro-pnictide materials is valid.
Whether magnetic moments in 
ferro-pnictide materials
are itinerant or localized is vigorously debated, however.
In particular,
the metallic temperature dependence shown by the electrical resistance
in the cSDW phase of ferro-pnictide materials
in addition to the absence of satellite peaks in inelastic X-ray scattering spectra 
is consistent with itinerant  magnetism.\cite{wang}$^,$\cite{yang_09}
The multi-orbital Heisenberg model Hamiltonian
(\ref{j0j1j2}) studied here
is also potentially consistent with the semi-metallic nature
of ferro-pnictide materials, on the other hand.  
Figure \ref{hidden_order} indicates that the classical hidden ferromagnetic state
remains intact in the presence of mobile holes 
that are prohibited from hopping between orbitals, for example.
Likewise, the classical hidden N\'eel state remains intact in the presence of mobile
holes that are restricted to hop between different orbitals at nearest-neighbor sites
of the square lattice of iron atoms.
Such limiting cases can result in low-energy electronic hole excitations near zero momentum,
which are consistent with the electronic structure revealed by
angle-resolved photoemission spectroscopy (ARPES) 
in the cSDW phase of ferro-pnictide materials.\cite{kondo_10}
Resonant scattering of such hole excitations
with the quantum-critical spin-waves studied here 
(Figs. \ref{sw} and \ref{cp_spctrm}) could also result 
in low-energy single-particle excitations
at momenta near those associated with cSDW order.
Recent ARPES  studies
 of BaFe$_2$As$_2$
in the cSDW phase reveal the existence of low-energy  electronic excitations
near such  momenta.\cite{richard_10}
The measured electronic spectrum there forms a Dirac cone,
with a Fermi velocity of about $330$ meV\,\AA.
It is an order of magnitude smaller than in graphene,
and it is hence consistent with strong electronic correlations.\cite{harrison_09}
It also remarkably  lies inside the range of  spin-wave velocities
extracted from low-energy magnetic excitations near  cSDW wave numbers
in the doped superconducting compound BaFe$_{2-x}$Co$_x$As$_2$:\cite {hayden_10} 
$v_t = 230$ meV\,\AA\ and $v_l = 580$ meV\,\AA\
(see Fig. \ref{qcp_fit}).
The above observations therefore indicate that 
injecting mobile holes into the states studied here
that display hidden magnetic order (Fig. \ref{hidden_order})
potentially can account for the nature of low-energy single-particle excitations
in ferro-pnictide materials.
They also
indicate that a description of the superconducting state of ferro-pnictide materials
in terms of a doped Mott insulator\cite{Si&A} is possible.

\begin{acknowledgments}
The author thanks Ed Rezayi, Radi Al Jishi and Elena Bascones for useful discussions.
Exact diagonalizations of the $J_0$-$J_1$-$J_2$ model (\ref{j0j1j2}) 
were carried out on the SGI Altix 4700 (Hawk)
at the AFRL DoD Supercomputer Resource Center.
This work was supported in part by the US Air Force
Office of Scientific Research under grant no. FA9550-09-1-0660.
\end{acknowledgments}

\end{document}